\documentclass{article}

\usepackage{PRIMEarxiv}
\usepackage{natbib}
\usepackage[utf8]{inputenc} % allow utf-8 input
\usepackage[T1]{fontenc}    % use 8-bit T1 fonts
\usepackage{hyperref}       % hyperlinks
\usepackage{url}            % simple URL typesetting
\usepackage{booktabs}       % professional-quality tables
\usepackage{amsfonts}       % blackboard math symbols
\usepackage{nicefrac}       % compact symbols for 1/2, etc.
\usepackage{microtype}      % microtypography
\usepackage{lipsum}
\usepackage{fancyhdr}       % header
\usepackage{graphicx}       % graphics
\graphicspath{{media/}}     % organize your images and other figures under media/ folder

\usepackage{subfloat}
\usepackage{tikz}
\usepackage{tikzscale}
\usetikzlibrary{shapes}
\usetikzlibrary{patterns}

\title{Exascale Workflow Applications and Middleware: An ExaWorks Retrospective}

\author{
  Aymen Alsaadi \\
  Electrical and Computer Engineering \\
  Rutgers University \\
  NJ, New Brunswick, USA\\
  \texttt{aymen.alsaadi@rutgers.edu} \\
  \And
  Mihael Hategan-Marandiuc \\
  Argonne National Laboratory \\
  IL, Lemont, USA \\
  University of Chicago\\
  Chicago, IL, USA\\
  \texttt{hategan@mcs.anl.gov} \\
  \And
  Ketan Maheshwari \\
  Oak Ridge National Laboratory \\
  Oak Ridge, TN, USA \\
  \texttt{maheshwarikc@ornl.gov} \\
  \And
  Andre Merzky \\
  Incomputable LLC \\
  Highland Park, NJ, USA \\
  \texttt{andremerzky@incomputable-systems.com}\\
  \And
  Mikhail Titov\\
  Brookhaven National Laboratory\\
  Upton, NY, USA\\
  \texttt{mtitov@bnl.gov}\\
  \And
  Matteo Turilli\\
  Brookhaven National Laboratory\\
  Upton, NY, USA\\
  Rutgers University \\
  NJ, New Brunswick, USA\\
  \texttt{mturilli@bnl.gov}\\
  \And
  Andreas Wilke\\
  Argonne National Laboratory \\
  IL, Lemont, USA \\
  \texttt{wilke@mcs.anl.gov} \\
  \And
  Justin M. Wozniak\\
  Argonne National Laboratory \\
  IL, Lemont, USA \\
  \texttt{wozniak@mcs.anl.gov} \\
  \And
  Kyle Chard\\
  Argonne National Laboratory \\
  IL, Lemont, USA \\
  University of Chicago\\
  Chicago, IL, USA\\
  \texttt{chard@uchicago.edu} \\
  \And
  Rafael Ferreira da Silva\\
  Oak Ridge National Laboratory\\
  Oak Ridge, TN, USA \\
  \texttt{silvarf@ornl.gov}
  \And
  Shantenu Jha\\
  Brookhaven National Laboratory\\
  Upton, NY, USA\\
  Rutgers University \\
  NJ, New Brunswick, USA\\
  \texttt{shantenu.jha@rutgers.edu}
  \And
  Daniel Laney\\
  Lawrence Livermore National Laboratories\\
  Livermore, CA, USA\\
  \texttt{laney1@llnl.gov}\\
}

\begin{document}
\maketitle

\begin{abstract}
Exascale computers offer transformative capabilities to combine data-driven and learning-based approaches with traditional simulation applications to accelerate scientific discovery and insight. However, these software combinations and integrations are difficult to achieve due to the challenges of coordinating and deploying heterogeneous software components on diverse and massive platforms. We present the ExaWorks project, which addresses many of these challenges. We developed a workflow Software Development Toolkit (SDK), a curated collection of workflow technologies 
that can be composed and interoperated through a common interface, engineered following current best practices, and specifically designed to work on HPC platforms. ExaWorks also developed  PSI/J, a job management abstraction API, to simplify the construction of portable software components and applications that can be used over various HPC schedulers.  The PSI/J API is a minimal interface for submitting and monitoring jobs and their execution state across multiple and commonly used HPC schedulers. We also describe several leading and innovative workflow examples of ExaWorks tools used on DOE leadership platforms. Furthermore, we discuss how our project is working with the workflow community, large computing facilities, and HPC platform vendors to address the requirements of workflows sustainably at the exascale.
\end{abstract}

% keywords can be removed

\section{Introduction}
\label{sec:introduction}
The benefits of high-performance workflows for scientific
discovery have been well-established~\cite {da2021community,badia2017workflows}. 
However, increasing heterogeneity of computing platforms, diversity of task types, and challenges associated with scaling will increase the prevalence and sophistication of high-performance workflows. Workflows are used to assemble multiple components (e.g., ModSim applications, meshing, post-processing, ML tools, custom scripts) into complex applications (e.g., ensembles, search, or optimization patterns) consisting of multiple stages, collected into longer-term studies, perhaps leveraging multiple compute resources and facilities~\citep{badia2017workflows, dasilva2024computer}. 

The unprecedented processing and analysis capabilities of 
high-performance computing (HPC) enables
scientists to tackle complex problems and simulations that were previously
infeasible. In particular, a key promise of such computers is the ability to
support sophisticated and scalable high-performance workflows as a vehicle to
accelerate scientific discovery across domains, from materials
development to climate science. In addition to enabling the simulation and
modeling of complex phenomena with unprecedented accuracy and detail,
high-performance workflows in conjunction with experimental and observational
facilities provide new insight and decision-making, hitherto deemed 
difficult~\citep{wcs2022}.

However, significant improvements in
existing capabilities and new computational methods will be required to capitalize on this potential fully. For example, advances in automation, novel real-time and streaming analysis, and coupling of AI methods to traditional HPC simulations are needed to
leverage increases in raw processing power. Furthermore, single and isolated
workflows will need to give way to campaigns that dynamically integrate
multiple adaptive workflows to work together and in collaboration. For example, recent work~\citep{saadi2021impeccable} illustrates the power of such
computational techniques for rapidly exploring and analyzing a vast
design space of therapeutics. Similar scenarios are
easily envisioned for climate and materials science, where discovering
novel materials with desirable properties requires multiple distinct methods,
algorithms, and workflows to operate in conjunction. In this era of
integrated computing and research facilities, breaking free of the ``tyranny'' of platforms designed merely for peak FLOPS instead of productive ones is imperative. 

This ``tyranny'' of computing platforms creates significant challenges for both scientists and computing facilities: (i)~it leads to an unsustainable proliferation of domain-specific workflow infrastructures that are tightly coupled to both their execution platform's software ecosystem and the domain responsible for their development~\citep{exaworks-survey, github-workflows}; (ii)~these tools must repeatedly implement complete functionality sets to meet application requirements, resulting in redundant development efforts; (iii)~the science of optimizing workflows for performance and productivity is poorly understood and intimately bound to specific tools rather than the workflow applications themselves; and (iv)~workflow applications must integrate multiple layers, combining workflow management tools with domain-specific capabilities.

Many scientists respond to these challenges by constructing workflow applications manually, leveraging system schedulers directly, and orchestrating via homegrown scripts.  This approach has the virtue of apparent simplicity: the scientist needs only assume that an HPC system has a batch scheduler to organize work and filesystems to manage data and enable orchestration. A well-recognized pattern~\citep{al2021exaworks} is that these bespoke workflow solutions become more challenging to maintain and scale as workflow requirements increase. 

The scientific community has identified a critical barrier: researchers often avoid integrated workflow management solutions due to their complexity and concerns about ``vendor lock-in"~\citep{wcs2022, ferreiradasilva2021works}. To address this challenge, the community must focus on making workflow technologies more modular and composable, allowing scientists to build complex workflow applications from simpler, interoperable components. This approach would provide flexibility while avoiding the maintenance burden of entirely custom solutions.

This paper describes how the ExaWorks project and its software capabilities address the socio-technical challenges of designing and developing high-performance workflow applications and middleware. ExaWorks provides community building blocks~\citep{turilli2019middleware} (e.g., ExaWorks SDK), common interfaces (e.g., PSI/J~\citep{hategan2023psi}), and community forums (e.g., workflow community summits). The following section outlines the primary types, characteristics, and challenges of workflow applications and middleware. The following section provides an overview of the history of the ExaWorks project. The subsequent sections focus on the core technical contributions of the ExaWorks SDK and PSI/J, followed by an extensive discussion of workflow applications that ExaWorks has impacted. We conclude by discussing open issues and future directions for exascale workflow applications, middleware development, and uptake.

\section{Workflow Applications and Challenges}

Our discussion will focus on four main types of high-performance workflow applications:
\begin{enumerate}
\item \textbf{Compute-intensive workflows} require interaction between interdependent simulation codes that run on different computing architectures. 
\item \textbf{AI-coupled workflows} involve the active coupling of AI methods to traditional HPC workflows;
\item \textbf{Workflows with Experiments in the loop;}
\item \textbf{Time-sensitive Workflows} that require real-time or end-to-end performance with high reliability
\end{enumerate}
These four types are not entirely disjoint but help us think about how workflow systems must evolve to support these motifs. Compute
intensive workflows often feature interaction or coupling of multiple simulation applications~\citep{turner2022exaam} but can also arise from data integration-intensive patterns that combine and analyze data from numerous sources.  AI-coupled workflows often use the AI/ML process to influence the progression of the HPC workflow; we distinguish these workflows because the AI process interacts with the workflow system directly to steer the overall process, requiring additional capabilities from the system. Optimization and control system approaches can also fall into this category.  Workflows with Experiments in the loop include elements of compute- and data-intensive applications but critically integrate experiments (active, passive, or event-driven) as part of the overall process. The experiments could be part of the inner or outer loop.  Finally, workflows with a time-sensitive pattern have urgency, requiring real-time or end-to-end performance with high reliability, such as for timely decision-making, experiment steering, and virtual proximity. Time-sensitive workflows can, in turn, belong to the first three categories. The needs of the diverse scientific workflow applications often lead to projects designing and developing their own, often ad-hoc, workflow solutions. 

We organize our discussion of the software and services around the following challenges and requirements:

\begin{enumerate}

\item Resource and data management: The allocation and scheduling of computational resources must be finely tuned to avoid bottlenecks and ensure the efficient execution of complex tasks, significantly as the system scales to accommodate larger workloads. Simultaneously, handling vast amounts of data demands robust data management strategies to ensure integrity, availability, and accessibility. Dealing with diverse data types and sources in a scalable HPC environment becomes complicated.

\item Distributed Orchestration: As HPC systems grow in complexity, orchestrating tasks across multiple distributed resources (including across facilities) requires meticulous coordination and synchronization. The challenges include managing dependencies between tasks, handling failures and exceptions, ensuring data consistency across different nodes (and sites), and optimizing resource utilization.  All of these require the ability to
exchange information between processes and workflow systems, and cross authentication boundaries in distributed workflows.

\item Planning and Scheduling: Besides supporting orchestration between tasks, workflows require scheduling different types of computation, with
diverse quality of service and temporal and spatial constraints. There might be a
need to dynamically spawn machine learning or simulation jobs for training or
inference. Workloads can grow and shrink over time, which traditional MPI
ranks might not be well-suited for. For example, training and uncertainty
quantification can necessitate many small jobs that must be
co-scheduled across various computational resources like CPUs, GPUs, and other
accelerators.

\item Dynamism: 
When complex orchestration is combined with planning and scheduling challenges, workflow systems must be designed to adapt swiftly and efficiently to changes in systems, technologies, data, and even objectives. We consider this an extra functionality, layered on top of orchestration and scheduling capabilities, of flexible algorithms and adaptive resource management that can respond to real-time changes without disrupting ongoing processes. Addressing the challenge of dynamism necessitates incorporating predictive modeling and adaptive controls to ensure that the HPC workflows can scale and evolve in harmony with an ever-changing landscape.

\item Complexity: Post-processing workflows dominate with technological advancements, especially in instruments that can generate
vast amounts of data. Retaining all
the data becomes untenable due to the sheer volume. There's a need for better
data management solutions, which might involve data reduction, projection into
fewer dimensions, compression, and real-time decision-making.

\item Customization and Portability: Each function and component in the workflow might need specialized mechanisms for monitoring and logging. The metrics to evaluate the performance and effectiveness of different elements can vary widely.
\end{enumerate}

Addressing these challenges systematically instead of locally and piecemeal will require new approaches. 

%-----------------------------------------
\section{The Road to ExaWorks}
\subsection{ExaWorks: The Genesis}

The ExaWorks project~\citep{al2021exaworks} was started in late 2019 to bring workflow support to the Exascale Computing Project (ECP) and foster community, enhance collaboration, and move towards modern software engineering methodologies.

It is instructive to consider the genesis of the ExaWorks project itself. The process was one of extensive socialization and a slow process of consensus building within ECP leadership and its advisors.  The first inklings of a workflow-focused project in ECP occurred as part of the ECP project's gap analysis conducted in early 2017.  This gap analysis was conducted across multiple areas in preparation for RFPs later that year. At the ECP annual meeting in January 2017, during the workflow breakout session, it was decided to produce a survey of application teams to understand workflow requirements. The survey was conducted later that spring, with the initial draft report released in April 2017.  The extensive survey investigated data speed and feeds, multi-application workflow needs, exascale hardware usage (including predictions of usage of as-yet-to-be-fielded storage systems), and other technical aspects of workflows.  
% The overall report was deemed interesting, but no compelling message came out of the survey. 
While the report provided an overview of ECP workflow needs, it did not include clear findings or recommendations on how to proceed.  
We assess that this was likely due to the complexity of the survey questions and the timing of the survey---we were asking newly formed teams what their workflow requirements would be in the seemingly distant future of 2021-2022. 

Even so, many of the eventual leaders of ExaWorks were already funded by ECP in various contexts, as applications teams in ECP were leveraging the set of technologies they represented.  Thus, discussions on the need for a workflow-focused project continued in the community at various meetings in 2017 and 2018.  Finally, in 2019, 
the topic of workflows in ECP was revisited, and an effort was undertaken to reassess the application teams' needs concerning their workflows. The ExaWorks leadership team was assembled that year, and discussions began on how a pilot effort could be conducted to review workflow needs and produce a report and possible proposal for a complete project.

The initial meeting of the PIs occurred at Supercomputing in November 2019. Phase I of ExaWorks, on restricted funding, was designed to run for six months to conduct 
a focused survey and to formulate a plan for a fully-funded workflows project in the ECP for fiscal year 2021.  A further study was constructed, this time shorter and more focused on user practice and pain points, and that survey resulted in a clearer picture of the opportunity cost of each ECP team building its workflow infrastructure. It is likely that by early 2020, most teams had a much better understanding of their needs and challenges as they prepared for the arrival of Exascale hardware. The survey report
\citep{Workflow_Survey_2017} showed that many bespoke workflows were planned or used, leveraging mainly shell scripts and Python tools. Building on this information, the ExaWorks team created a plan and proposal for a fully scoped project later that year.

The ExaWorks project and its leadership team represented an ideal outcome in many ways. The leadership team was chosen among workflow projects that already had some level of adoption by one or more ECP application team (both Office of Science and NNSA), had proven scalability, and a high level of software engineering.  The leadership team established an early culture of collaboration, cooperation, and an avoidance of zero-sum thinking. This approach was vital in building a coalition of partners and stakeholders that could enable the eventual creation of the ExaWorks software development kit (SDK) and the cross-pollination of workflow technologies that occurred later in the project.

\subsection{ECP Applications Survey}
We surveyed ECP application teams in Phase I of ExaWorks to understand their workflow requirements and challenges.  The survey was conducted in two parts: an online questionnaire and targeted deep-dive interviews with a subset of teams.  We summarize here the results from this survey~\citep{exaworks-survey}. 

We sent the online questionnaire to 24 ECP applications teams and the 5 ECP co-design centers.  We received responses from 15 of the 29 teams. After reviewing these responses, we identified five teams to interview in depth.  Our selection criteria emphasized teams developing workflows that had either written or leveraged workflow management tools.  Our goal was to broaden our understanding of these workflows and the tools these teams employ.
Responses to the survey highlighted that many ECP application teams were orchestrating workflows using homegrown scripts (shell, Python, Perl) and tools like Make. Some teams reported using workflow tools:  Airflow, Cheetah, Fireworks, libEnsemble, Merlin, Nexus, Parsl, and Savannah.  Note that we allowed respondents to define ``workflow tool'' broadly, resulting in a mixture of general workflow tools and tools under development for particular sub-domains in HPC.  

We asked teams to describe their workflows. Using these descriptions, we grouped responses into the following motifs: 
\begin{enumerate}
\item \textbf{Single simulations:} workflows managing a single simulation, composed of various independent tasks and often scaling to extreme scale.
\item \textbf{Ensembles:} sets of runs, often statically defined parameter studies, parameter sweeps, and convergence studies.
\item \textbf{Analysis:} experiment-driven workflows involving a mixture of short/small jobs and larger analysis jobs.
\item \textbf{Dynamic:} workflows in which the runs are unknown \textit{a priori} and involve co-scheduling disparate tasks and orchestration among tasks. Integrated HPC and Machine Learning workflows are a growing and essential example.
\end{enumerate}

The ensemble motif was the most common motif reported by survey respondents, and often, these ensembles were managed via bespoke scripts. 
While one might expect that single simulations would be more common, it is likely that these teams did not employ workflow systems and were thus 
less likely to respond to the survey.  Analysis and AI/ML workflows featured in several responses, and even when these teams employed general-purpose workflow management systems, we found that a significant amount of customized internally developed infrastructure was still required.

We asked respondents to describe the following aspects of their workflows, and we summarize the responses here;
\begin{enumerate}
\item \textbf{Internal Orchestration:}
We aimed to understand the need for tasks in a workflow or single batch job allocation to interact with one another. 
Responses indicated the use of such coordination but limited communication
between tasks \textemdash{} though one responding team utilizes
streaming/service-oriented workflows where task-to-task interaction was
required.

\item \textbf{External Orchestration:}
We aimed to understand the extent to which teams utilized multiple machines or executed workflows across numerous machines.  The responses were evenly divided, with about half of the respondents indicating that their workflows span systems or that they would run them in that mode if they had a workflow tool that makes it possible to do so.  In most cases, the use of multiple systems was driven by the need to scale workloads and reduce computation time rather than a differentiation based on hardware or data locality.  Some teams described workloads that exceed scheduler job time limits, requiring submission of several batch jobs, 
and they considered this to be an external orchestration.

\item \textbf{Homogeneous vs. Heterogeneous tasks:}
In general, most respondents indicated a large dynamic range of job sizes.  Reasons for this range include scaling/convergence studies, simulation vs. analysis jobs, and co-scheduling of ML and simulation tasks.  Unsurprisingly, we found that teams with more complex and dynamic workflows reported high levels of task heterogeneity.
\end{enumerate}

The responses and our interviews with teams provided a strong finding that
supporting complex dynamic workflows across multiple machines/data centers
and porting to new machines is expensive in terms of developer time.  Each
cluster, even those that outwardly appear similar (e.g., Linux OS, Slurm batch
scheduler, etc.), requires customization in the workflow.  The subset of ECP
projects that need to run at multiple facilities have developed independent abstraction layers to support these customizations.   A key takeaway is that attacking the workflow management stack's lower layers can increase portability and reduce team costs. Finally, a common theme running through the survey is that developing robust, fault-tolerant and portable workflows is both a pain point and often a determining factor in whether a team will adopt a third-party workflow technology rather than creating their bespoke capability.  

The aforementioned salient points informed the scope, priorities, and approach of the ExaWorks project (e.g., SDK, PSI/J portability layer for schedulers), which we now discuss.
%-----------------------------------------
\section{ExaWorks SDK}
ExaWorks SDK is a curated collection of workflow technologies. The curation process includes (1) defining the criteria to select the technologies to be included in the SDK; (2) policies to ensure that those technologies are maintainable long-term; (3) analysis of the technologies design and implementation to assess whether they can be integrated to deliver new capabilities; (4) packaging suitable to be deployed on DOE HPC platforms; (5) continuous integration to test each technology on a growing number of DOE HPC platforms; and (6) documentation specifically designed to support demos, tutorials, hackathons but also end-to-end testing of exemplar use cases. 

ExaWorks SDK seeding technologies were chosen based on the requirements elicited by engaging with the ECP workflow communities. In 2017, a survey of workflow needs was performed~\citep{Workflow_Survey_2017}.  We created a survey to identify existing workflow systems efforts, both ECP-related and within the broader DOE software ecosystem.  That survey helped us understand the challenges, needs, and possible collaborative opportunities between workflow systems and the ExaWorks project. We took a broad view of workflows, including automated orchestration of complex tasks on HPC systems (e.g., DAG-based and job packing), coupling simulations at different scales, adaptive/dynamic machine learning applications, and other efforts in which a variety of possibly related tasks have to be executed at scale on HPC platforms. For example, after eliciting a brief description of an exemplar workflow, we asked about internal/external workflow coordination, task homogeneity/heterogeneity, and details about the adopted workflow tools.

Alongside the outcome of more than twelve community meetings, the results of our survey highlighted the state of the art of scientific workflows in the ECP community. To summarize, many teams are creating infrastructures to couple multiple applications, manage jobs---sometimes dynamically---and  
orchestrate compute/analysis tasks within a single workflow and manage data staging within and outside the HPC platforms. Overall, there is an evident duplication of effort in developing and maintaining infrastructures with similar capabilities. Further, customized workflow tools incur significant costs to port, maintain, and scale bespoke solutions that serve single-use cases on specific platforms and resources. These tools do not always interface with facilities smoothly and are difficult and costly to port across facilities. Finally, the lack of proper software engineering methodologies leads to repeated failures, difficulty in debugging, and expensive fixes. Overall, there was agreement in the community that costs could be minimized, and quality could be improved by creating a reliable, scalable, portable SDK for workflows. 

Based on the requirements summarized above, the ExaWorks SDK collects software components that enable the execution of scientific workflows on HPC platforms. We consider a reference stack that allows the development of scientific workflow applications, resolving the task dependencies of that application, acquiring resources for executing those tasks on a target HPC platform, and then managing the execution of the tasks on those resources. The SDK includes components that deliver a well-defined class of capabilities of that reference stack with different user-facing interfaces and runtime capabilities. Consistently, the SDK is designed to grow its components, including software systems that are open source (i.e., released under a permissive or copyleft license) and developed according to software engineering best practices, such as test-driven development, release early and often, and version control.

\subsection{Components and Integration}

\begin{figure}[ht]
     \centering
     \includegraphics[width=1\textwidth]{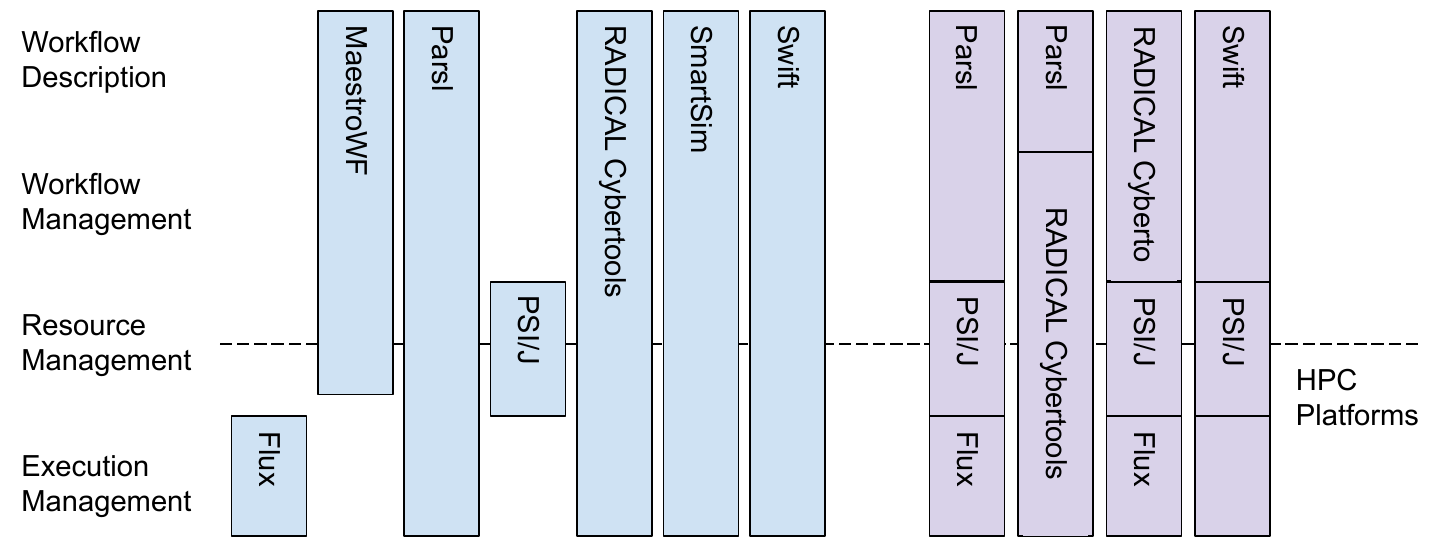}
     \caption{The ExaWorks SDK provides various interfaces, programming models, and runtime capabilities for executing scientific workflows at scale on high-performance computing (HPC) platforms. The reference stack (represented by the blue boxes) includes components that enable end-to-end capabilities, from workflow description to execution management (indicated by the vertical extent). The purple boxes highlight the integrations of multiple components, achieved by having each component expose its APIs for workflow and resource management, among other functions. For instance, the left-most purple stack illustrates the integration of PSI/J, which replaces Parsl's built-in scheduler support, alongside the use of Flux for job scheduling. The dashed line signifies the boundary between the workflow systems (operating in user space) and the specific capabilities of HPC platforms (such as Slurm). Both Flux and PSI/J operate in user and system space, with PSI/J bridging these two areas. Importantly, both systems remain independent of any system-specific plug-ins.}
     \label{fig:sdk_stack}
\end{figure}

The SDK contains seven software components: Flux~\citep{flux, AHN2020202}, MaestroWF~\citep{dinatale19maestro}, Parsl~\citep{babuji19parsl}, PSI/J~\citep{hategan2023psi}, RADICAL Cybertools~\citep{balasubramanian2016ensemble,merzky2021design}, SmartSim~\citep{smartsim}, and Swift/T~\citep{wozniak2013swift}. As shown in Fig~\ref{fig:sdk_stack}, each component delivers capabilities end-to-end or at a specific level of our reference workflow stack. That way, the SDK offers users alternative tools across or along the reference stack, depending on their specific requirements. For example, Parsl, RADICAL Cybertools, SmartSIM, Swift/T, and to a certain extent, MaestroWF offer end-to-end workflow capabilities but with different programming models, application programming interfaces (API), support for HPC platforms, and scientific domain. Flux and PSI/J offer capabilities focused on resource and execution management, enabling the execution of user-defined workflows on various HPC platforms. 

Further, we have encouraged each component to expose well-defined interfaces for sub-components. 
%, exposing well-defined interfaces. 
That allows us to promote integration among components, obtain new capabilities, and avoid lock-in to specific solutions maintained by a single team or designed to support a particular class of scientific problems and platforms. The underlying idea is to prevent reimplementing existing capabilities, instead investing in designing and coding integration layers between the technologies with diverse capabilities. While the concept is simple, its actual implementation is challenging. Software systems developed by independent engineering teams are not thought to be compatible at the abstraction and implementation levels. 

Our integration experience revealed a consistency of abstractions in software designed to support scientific workflows. For example, most tools share analogous task abstractions, assumptions about data dependencies among tasks, and, to some extent, the internal representation of computing resources and how they relate to computing tasks. Most differences were found at the interface and runtime level, where each tool implements distinctive designs and capabilities. SDK tools adopt similar best engineering practices (as most software is designed for production these days), contributing to a certain degree of homogeneity. For example, most SDK tools adopt designs based on well-defined APIs, connectors, and adaptors. Finally, implementation-wise, some tools adopted similar programming languages and technologies, such as Python and ZeroMQ. 

Integration points were clearly identified at the level of connectors and adapters, focusing primarily on translating internal representations of tasks, data, and resource requirements. We can integrate user-facing capabilities with various runtime features, requiring minimal code to be written, primarily to implement translation layers among well-defined interfaces or connectors from existing base classes. Integrating models of resource representations presented significant challenges, particularly in relation to resource acquisition capabilities. At times, this required enhancing existing user interfaces to create a unified definition of resource requirements. Other instances necessitated adjustments to the data structures and methods used to manage resources during runtime. Despite these code extensions and modifications, the overall integration process was relatively straightforward from both design and implementation standpoints.

The experience gained from developing the SDK demonstrates that tools should be designed following best engineering practices to mitigate and reduce fragmentation in the landscape of scientific workflow technologies. This includes using established patterns for distributed systems, adopting common abstractions, and applying consistent patterns along with separating concerns for communication and coordination. By implementing these strategies, tools are more likely to be compatible with future third-party applications.

In Fig.~\ref{fig:sdk_stack}, the purple boxes illustrate how we integrated various components: Parsl with RADICAL-Pilot, Parsl with PSI/J, RADICAL-Pilot with Flux, RADICAL-Pilot with PSI/J, and Swift/T with PSI/J. Each integration enhances the capabilities at different levels of the end-to-end stack necessary for executing workflow applications on HPC platforms.

Specifically, the integration of RADICAL-Pilot with Parsl enables the effective and efficient execution of MPI tasks, which can be implemented as either standalone executables or Python functions, at scale on Department of Energy (DOE) platforms, including the exascale leadership-class Frontier machine. PSI/J provides application portability for Parsl, RADICAL-Pilot, and Swift/T, allowing users to run applications across most DOE HPC platforms seamlessly.

Additionally, integrating Flux with Parsl and RADICAL-Pilot offers advanced scheduling and launching capabilities that would otherwise be difficult to achieve with each system's standalone launch methods. Importantly, all these integrations are isolated within each tool, allowing for flexible compositions based on specific use case requirements. For example, users can create combinations such as Parsl + RADICAL-Pilot + Flux, Parsl + Flux, or RADICAL-Pilot + PSI/J, among others.

\subsection{Testing and Documentation}

The ExaWorks SDK solves the testing challenge by offering an infrastructure that facilitates the testing of SDK components on DOE platforms as well as other accessible platforms. This infrastructure includes a test runner framework and a dashboard.

The test runner framework comprises a set of tools and practices designed to facilitate the deployment of SDK components and the execution of tests. The primary driving force behind this framework is the GitLab Continuous Integration (CI) pipelines utilized at various Department of Energy (DOE) laboratories. Additionally, support for GitHub Actions and direct invocation—via the \texttt{cron} tool—is also included.

Each pipeline consists of two main steps: deployment and test execution. Three deployment methods are available: \texttt{pip}, \texttt{conda}, and \texttt{spack}, although support for these methods may vary depending on the package being tested and the environment where the pipeline is executed. A location-agnostic version of the SDK is also accessible as a \texttt{Docker} container.

Active pipelines are configured for the Argonne Leadership Computing Facility (ALCF) at Argonne National Laboratory (Polaris), Lawrence Livermore National Laboratory (LLNL) (Lassen, Quartz, and Ruby), the National Energy Research Scientific Computing Center (NERSC) (Perlmutter), and Oak Ridge Leadership Computing Facility (OLCF) at Oak Ridge National Laboratory (Ascent and Summit). 

The test execution step runs a series of validation tests for each SDK component and integration tests that verify the correct interaction between two or more SDK components when applicable.

A comprehensive testing infrastructure requires effective methods for collecting and presenting results to both developers and potential users. The ExaWorks SDK testing dashboard fulfills this essential role by providing a mechanism to report test results to developers and, in its final production version, to users as well. This dashboard was initially created for the PSI/J-Python project, where it facilitated the centralization of results from user-maintained test runs. Similar to PSI/J-Python, the ExaWorks SDK employs a hybrid approach: some tests are managed by the ExaWorks team on specific HPC systems, while users are empowered to run tests on machines they control.

The testing process is outlined as follows: A client-side test runner, typically triggered by a GitLab pipeline, executes the desired tests and captures the results along with additional information, such as output streams and logs. This information is then uploaded to the testing dashboard, where it is presented to users and developers. 

By default, the results are aggregated by site and date in a calendar view, providing a quick overview of overall pass/fail trends over the past few days. Users can select specific sites and navigate to individual test runs, allowing them to examine detailed test results, as well as client-provided outputs and logs.

The testing dashboard is composed of two main components: a backend and a frontend. The backend handles user authentication, stores test results in a database and responds to queries for historical test data. The frontend is a web application built using the Vue.js~\citep{vuejs-web} library, which displays test information for both developers and users.

Similar to the PSI/J-Python testing dashboard, the SDK dashboard employs a straightforward authentication mechanism that requires users to verify their email addresses. This process ensures that result uploads are associated with a specific identity, which helps manage access to the dashboard. The decision to use simple email validation instead of a more complex authentication provider service is driven by the aim to accommodate test results from users across various institutions and private individuals.

Alongside testing, documentation is also a fundamental element of the ExaWorks SDK. SDK documentation has to satisfy three main requirements: (i)~centralize into a single venue and under a consistent interface all the information specific to SDK; (ii)~avoid duplication of documentation between SDK and its tools; (iii)~minimize maintenance overheads; (iv)~manage a rapid rate of obsolescence in the presence of continuously and independently updated software components; (v)~use the same documentation for multiple purposes like training, dissemination, hackathons, and tutorials.

Documenting the SDK poses specific challenges compared to documenting a single software system. The SDK is a collection of software components independently developed by unrelated development teams. While each tool is part of the SDK, the ExaWorks team does not participate in the development activities supporting each component, decide when and how those components are released, or update or extend each component's documentation. That has the potential to make the SDK documentation obsolete, very resource-intensive to maintain, and prone to create duplication detrimental to the end user.

We consistently scoped the SDK documentation to offer static and dynamic information. Static information focuses on the SDK itself, offering the needed information about what it is and it is not, the list of current components, the minimum requisites for an element to be part of the SDK, the process to follow to include that component, and details about code of conduct and governance. Due to the above-mentioned challenges, we avoid static information about the SDK components and directly link specific documentation for each tool. As such, we provide a hub where users can find a variety of pointers to the vast and often dispersed software ecosystem to support the execution of scientific workflows at scale on DOE HPC platforms. 

We devised a novel approach to dynamic documentation centered on tutorials designed to be used for outreach, training, and hackathon events. At the core of our strategy are Jupyter notebooks containing both documentation and code to deliver: (i)~paradigmatic examples of scientific applications developed with the SDK components and executed on DOE platforms; (ii)~tutorials about capabilities of SDK that serve the specific requirements of the DOE workflow community; (iii)~detailed examples of resource acquisition, management, tasks definition and scheduling; (iv)~debugging and tracing workflow executions; and (v)~many other common tasks required by executing workflows on HPC platforms.

Organizing the dynamic component of SDK documentation on GitHub allowed us to extend its use beyond its traditional boundaries. Utilizing GitHub workflows, we created a continuous integration platform where each Jupyter notebook can be automatically executed every time any SDK component is released. That avoids manually maintaining all the tutorials and makes it immediately apparent when a new component release breaks the tutorial's code. Further, the same tutorials can be seamlessly integrated with Readthedocs~\citep{sdk-readthedocs}, the system we use to compile and distribute the SDK documentation. Executing the tutorials every time the documentation is published guarantees that the tutorials work, significantly improving the quality of the information distributed to the end user. Finally, as part of the GitHub workflows, we package all the tutorials into a Docker container~\citep{sdk-container} that enables users to execute the tutorials both locally or via Binder~\citep{binder-web} with minimal overheads and portability issues.

\subsection{Experiences using multiple SDK Components}

We worked to integrate workflow components and apply them to various applications.  We report on two examples highlighting the benefits of 
common interfaces enabling integration.

\subsubsection{Parsl + WQ for DESC.}

DESC (Dark Energy Science Collaboration) is focused on  
understanding the nature of dark energy by analyzing
vast volumes of data obtained from the Rubin Observatory. 
The collaboration extensively uses workflows to process
survey data, conduct user analyses, and create simulation 
data. They run workflows at large scales, up to thousands
of nodes on various supercomputers. 
One workflow, \textit{ctrl\_bps}, designed to support
LSST Batch Production Service (BPS) is used to process
survey images. DESC has implemented a Parsl~\citep{babuji19parsl} plugin
to run the workflow on HPC systems.

In this particular workflow, processing can be
parallelized across images. However, 
the compute requirements for each patch are variable
and thus, it is challenging to allocate work efficiently 
resources---too small and resources are underutilized, too 
large, and computations fail. 
To address this need, we integrated Parsl and WorkQueue via a common interface (the Python concurrent futures executor
API). The Parsl-based workflow passes
work (in Python tasks) to WorkQueue to 
execute on provisioned resources. WorkQueue can determine task resource requirements dynamically. 
Thus, by integrating these two tools, users benefit from
the same Parsl interface, DAG processing, data management, etc.
While also exploiting WorkQueue's resource management abilities.

\subsubsection{Parsl + RADICAL Cybertools for quantum chemistry}

There are myriad simulation methods for quantum chemistry
calculations, ranging from short-running, single-core 
calculations through to long-running, multi-node MPI-based
calculations. 
Colmena~\citep{wardColmena2021} is an open-source Python framework designed to enable
the steering of computational campaigns composed of
repeated runs of multi-scale simulation codes. Colmena
builds on Parsl to manage the execution of tasks on resources.
While Colmena has been used for various applications, 
it has focused primarily on single-node simulations, machine-learning training, and inference. 

Applying the component-based approach described previously, we
extended Colmena to run a broader range of tasks, including
variable-shaped MPI simulation by integrating RADICAL-Pilot~\citep{alsaadiParslRP2022}. 
Specifically, we implemented a new Parsl executor, RPEX, 
by implementing the defined interface. RPEX enhances Colmena’s capabilities by concurrently distributing and executing MPI Python functions alongside (non)MPI executable tasks. This integration significantly broadens the range of simulations Colmena can handle. The integration required minimal modifications to RP and Parsl. The most substantial change involved updating the Parsl configuration and task resource specifications to accommodate MPI tasks.
%------------------------------------------
\section{Portable Submission Interface for Jobs (PSI/J)}
%!TEX root = ./main.tex

PSI/J is a job management abstraction API that simplifies the construction of portable software components and applications over various HPC schedulers.  The PSI/J API is designed as a minimal interface for submitting and monitoring jobs and their execution state. The need for minimalism is informed, in part, by the observations in the introduction that complexity is unlikely to lead to a successful job abstraction API solution in the long term. That is coupled with the observation that many custom solutions are focused almost exclusively on submitting and monitoring jobs with no further adornments.

The PSI/J API is also designed to allow scalable implementations where scalability is targeted in the number of handled jobs and the rate of job submission. Our reference Python PSI/J implementation fully uses the API's scalability features. Specifically, the API is asynchronous to support threadless use. The choice of asynchronous API is also motivated by the fact that one can quickly transform an asynchronous API into a synchronous one without incurring a performance penalty. 

In PSI/J, the simplicity of the API is favored over that of the implementation if a reasonable implementation choice exists for a given design goal~\citep{den2010price}. This is motivated by the fact that implementation complexity can be addressed with reusable solutions that multiple implementations can share. In contrast, API complexity translates into a pervasive overhead for API users. For example, bulk operations can, in some instances, improve the scalability of an implementation. Bulk operations are versions of API operations that act on multiple items instead of single ones. A bulk \emph{submit} method would take a list of jobs as arguments and submit them all in one call to the underlying local resource manager (LRM), assuming that the alternative of making multiple calls to the LRM introduces a significant combined overhead. However, a simple time windowing function can aggregate individual job submissions and transparently route the resulting list of jobs to a bulk LRM submit call if such a call is available. This would retain the simplicity of the API while also allowing the implementation to be efficient, which is the favored choice. Certain caveats of this approach should, however, be noted. Time-based clustering of jobs is insufficient in determining jobs related to their various properties and may be unsuitable in creating job arrays as supported by multiple LRMs. Such support may be added to PSI/J in the future.

A similar problem to bulk submission is bulk job status querying. Invoking \emph{qstat} commands individually for each job can result in poor scaling as the number of actively managed jobs increases. The PSI/J specification~\citep{psijspec} mandates that implementations query the status of jobs in bulk to avoid this issue.

The PSI/J specification is organized into three primary layers, depending on the level of functionality that it describes. The \emph{local layer} (see Fig.~\ref{layer0}) defines the API needed to interact with job schedulers locally. That is, the location of the job scheduler is implicit and assumed to be on the same machine as the one on which the client application is running. The \emph{remote layer} (see Fig.~\ref{layer1}) defines additional API elements needed to submit jobs to a remote scheduler running on a different machine than the one on which client code is running. The \emph{nested layer} (Layer 2) adds API elements needed to interact with pilot job implementations. At the time of writing this paper, only the local layer of the PSI/J specification is available publicly, and work is underway to draft the remote layer.

\newcommand{\cbox}[3]{
    \draw[rounded corners, fill=white] (#1) rectangle (#2)
        node[pos=0.5, align=center] {#3};
}

\tikzset{>=latex}
\tikzset{
    every picture/.style={line width=0.4pt}
}

\def\layerzero{
    \cbox{-0.2, 1.5}{8, 3.0}{PSI/J Core};

    \cbox{0, -1.2}{2, -0.2}{Slurm\\LRM}
    \cbox{2.2, -1.2}{4.2, -0.2}{PBS\\LRM}
    \cbox{4.4, -1.2}{6.6, -0.2}{LSF\\LRM}

    \draw[<->] (1, -0.2) -- (1, 1);
    \draw[<->] (3.2, -0.2) -- (3.2, 1);
    \draw[<->] (5.4, -0.2) -- (5.4, 1);

    \cbox{0, 1}{2, 2.0}{Slurm\\Executor}
    \cbox{2.2, 1}{4.2, 2.0}{PBS\\Executor}
    \cbox{4.4, 1}{6.6, 2.0}{LSF\\Executor}

    \draw (7.4, 1.1) node {\Large$\cdots$};
}
\begin{figure}[htb]
    \centering
    \begin{minipage}{0.48\textwidth}
        \centering
        \begin{tikzpicture}[font=\sffamily\footnotesize, scale=0.68]
            \layerzero
            \draw [ultra thin, gray, dashed] (-0.5, 0.4) -- (10, 0.4);
            \node[align=center] at (9, -0.2) {System\\Software};
            \node[align=center] at (9, 1) {User\\Software};
        \end{tikzpicture}
        \vspace{2.9cm}
        \caption{Illustration of the local layer of PSI/J.}
        \label{layer0}
    \end{minipage}\hfill
    \begin{minipage}{0.48\textwidth}
        \centering
        \begin{tikzpicture}[font=\sffamily\footnotesize, scale=0.78]
            \begin{scope}[shift={(0.15, -3.8)}, scale=0.7, font=\sffamily\scriptsize]
                \cbox{-0.2, 3.6}{8, 5.1}{PSI/J Service}
                \cbox{8.3, 3.6}{10.7, 5.1}{SSH\\Service}
                \layerzero
                \draw[<->] (3.9, 3.6) -- (3.9, 3.0);
            \end{scope}

            \draw [ultra thin, gray, dashed] (-0.5, 0.4) -- (9.5, 0.4);
            \cbox{-0.2, 1.5}{8, 3.0}{PSI/J Core};
            \cbox{0, 1}{5.8, 2.0}{Remote Invocation Executor};
            \cbox{6.0, 1}{7.65, 2.0}{SSH\\Executor};

            \draw[<->] (2.9, -0.2) -- (2.9, 1);
            \draw[<->] (6.825, -0.2) -- (6.825, 1);

            \draw (7.4, 2.3) node {\Large$\cdots$};

            \node[align=center] at (9, 0) {Remote};
            \node[align=center] at (9, 0.8) {Local};
        \end{tikzpicture}
        \caption{Intended usage scenario for the remote layer of PSI/J.}    
        \label{layer1}
    \end{minipage}
\end{figure}

In the remote layer, multiple remote schedulers on multiple HPC schedulers are meant to be accessible concurrently from the same client process. Similarly, in the local layer, it is desirable to submit simple test jobs that can be run locally using a forked process. To support this scenario, PSI/J-Python adopts a multiple-dispatch mechanism in which bindings to underlying job execution mechanisms can coexist and be used concurrently. Such bindings are called ``executors''. In this sense, it differs fundamentally from most DRMAA implementations, which require the client executable to be dynamically or statically linked to switch to an alternate LRM.

The Python implementation of PSI/J uses a dynamic plugin discovery mechanism that allows a PSI/J core to detect executor and launcher implementations installed in different places from the PSI/J core. This allows for various scenarios, such as a stable system-provided core using a user-customized executor implementation or a user-installed core using a system-provided executor implementation (similar to DRMAA). Executors for Slurm, PBSPro, LSF, Cobalt, and Flux~\citep{flux} are provided with the current version of the reference implementation. A ``local'' executor that runs jobs using a simple fork mechanism is also provided. Launcher implementations are provided for all LRM-specific launchers (\emph{srun}, \emph{aprun}, \emph{jsrun}, etc.) as well as for generic launchers, such as \emph{mpirun}. The dynamic plugin system allows PSI/J Python to be extensible with new executors and launchers without necessarily requiring that the additional executors or launchers be part of the PSI/J Python code base. 

We have opted for the executors provided by the reference PSI/J Python implementation to use publicly available LRM interfaces, which usually consist of well-known commands, such as \textit{qsub} and \textit{qstat}. This deliberate choice assumes that established public LRM interfaces are less likely to change and lead to incompatibilities than proprietary ones. This contrasts with many DRMAA implementations, which use proprietary APIs to communicate with LRMs. However, the choice made by the PSI/J Python reference implementation does not preclude one from writing executors using proprietary APIs.
\subsection{PSI/J and Facilities}

PSI/J~\citep{hategan2023psi} is a language-independent API that addresses a small but fundamental problem in HPC: uniform access to batch schedulers and other job execution mechanisms. It provides an abstraction layer over the different ways of submitting and monitoring jobs. This problem is widespread across higher-level software that aims for portability across multiple HPC systems. It is almost universally addressed ad-hoc, which results in significant work duplication and other undesirable aspects. The PSI/J API comes with a reference Python implementation. Python was chosen due to its extensive use in workflow, experiment management, and other high-level tools. Learning from several similar past projects~\citep{hategan2023psi}, we welcome contributions from the community in all aspects. It also encourages users to make such contributions using several strategies:
\begin{itemize}
 \item Supporting user-space deployments allows users to test and use their modifications instantly.
 \item Allowing user plugins to be used with system deployments enables a hybrid approach.
 \item Implementing a testing infrastructure that allows users to run tests on their infrastructure and submit results to a central location that developers can use to address issues proactively.
 \end{itemize}

In advancing multi-facilities integration, the ExaWorks project has demonstrated a paradigm shift by adopting PSI/J across various facilities, epitomizing seamless interoperability amidst diverse and heterogeneous systems. For instance, at ORNL, the INTERSECT~\citep{engelmann22intersect} initiative is leveraging PSI/J to enable science use cases that span an intricate tapestry of computational resources. PSI/J, by design, bridges the gap between the complex specifics of individual resource managers and the overarching need for a unified, flexible computing environment (e.g., OLCF's Advanced Computing Ecosystem (ACE) IRI testbed). 
%------------------------------------------
\section{Workflow Applications and Science Impact}

\subsection{SC20 ACM Gordon Bell Special Award}

The winner and two of three finalists of the SC20 Gordon Bell Special Award
for COVID-19 competition leveraged ExaWorks technologies. 
This is a demonstration of the effectiveness of using high-quality, scalable workflow building blocks to create sophisticated dynamic workflows that can leverage leadership-class supercomputers. 
All four COVID-19 award finalists involved workflows; three used ExaWorks technologies. Each team developed tailored workflow solutions, leveraging ExaWorks technologies at crucial points, to enable scalability while reducing the developer time needed to support the scale and magnitude of these research efforts.

For example, the award's winner addressed the challenge of evaluating a potentially huge set of ``biologically interesting'' conformational changes by creating ``a generalizable AI-driven workflow that leverages heterogeneous HPC
resources to explore the time-dependent dynamics of molecular systems.'' This workflow used DeepDriveMD and components from the ExaWorks SDK. It combined
cutting-edge AI techniques with the highly scalable NAMD code to produce a new
high watermark for classical MD simulation of viruses by simulating 305 million atoms. The ORNL Summit system was able to deliver impressive, sustained
NAMD simulation performance, parallel speedup, and scaling efficiency for the
complete SARS-CoV-2 virion. AI helped identify interesting conformational changes
explored further to understand the critical molecular changes due to the ``jiggling and wiggling of atoms.''

Another finalist used Flux to provide the scalable backbone of Livermore’s Rapid COVID-19 Small Molecule Drug Design workflow. Flux provided an unprecedented level of workflow system composability, enabling highly complex campaigns such as drug design to be efficiently architected promptly.

The third finalist adopted Swift/T to develop a highly scalable epidemiological simulation and machine learning (ML) platform. The workflow was a complex structure of CityCOVID, a parallel RepastHPC agent-based modeling simulation of the 2.7 million residents of Chicago interspersed with series of ML-accelerated optimization tasks.
Multiple Swift/T design features aided in integrating the complete CityCOVID and ML epidemiological modeling platform. The simulation is a stand-alone C++ module that, in this case, ran on 256 cores and communicated internally with MPI. Invoking large, concurrent batches of such runs efficiently is one of the main capabilities of the Swift/T runtime, which invokes the simulator repeatedly through library interfaces.

Another challenge was generating and coordinating many single-node optimization tasks using vendor-optimized multithreaded math kernels. These single-node tasks were calls to various R libraries, dispatched via a custom R parallel backend. This extended the notion of workflow composability, a vital theme of the ExaWorks project, into the algorithmic control of the simulation and learning through external algorithms developed in ``native'' ML languages R and Python. This was implemented using the resident or stateful task capabilities of Swift/T and the associated EMEWS algorithm control framework.

\subsection{Genome-scale language models}
The 2023 Gordon Bell special prize for COVID research was awarded to a multi-institution team focused on studying how new and emergent variants of pandemic-causing viruses, specifically SARS-CoV-2, are identified and classified.  The team took a unique approach by adapting large language models (LLMs) for genomic data to create genome-scale language models (GenSLMs) that can learn the evolutionary landscape of SARS-CoV-2 genomes~\citep{chow23genslms}.
The team pre-trained their model on 110 million prokaryotic gene sequences and then fine-tuned a SARS-CoV-2 model on 1.5 million genomes. Their results showed that GenSLMs can accurately identify variants of concern.  GenSLMs were trained using two GPU-based supercomputers. GenSLMs uses Colmena~\citep{wardColmena2021}, a Python framework for steering campaigns, built on top of Parsl~\citep{babuji19parsl}. Specifically, it uses this workflow architecture with the trained model to design new proteins. 

\subsection{Materials design}
The ExaLearn project used Colmena~\citep{wardColmena2021} (and Parsl) in various applications. One example is in the design of molecules for redox-flow batteries. Specifically, the AI-based workflow runs various tasks that 1) compute the performance of a molecule (i.e., solvation energy, redox potential), 2) train models to predict molecule performance, and 3) use those models to infer the performance of new molecules. The approach delivered a 20\% increase in the number of high-performing molecules identified. The workflow was tuned to efficiently use HPC resources, for example, by overlaying communication, exploiting specialized hardware, and efficiently scheduling the different workload components. Further use of this framework in protein design and molecular dynamics is outlined in~\citep{colmenaecp}.

\subsection{Exascale Additive Manufacturing (ExaAM)} The rich variety of ExaWorks tools, along with their integration capabilities, facilitated the start of a strong collaboration with the ECP ExaAM (Exascale Additive Manufacturing) project. The ExaAM team has developed a suite of exascale-ready computational tools to model the process-to-structure-to-properties (PSP) relationship for additively manufactured metal components. Their uncertainty quantification (UQ) workflow targets to quantify the effect that uncertainty has on local mechanical responses in processing conditions. In their workflow, the management of single simulation runs didn't require sophisticated processes, but to coordinate 7875 simulation runs, it became crucial to look for an ensemble management system. Essential requirements to the management system were: (i) the ability to coordinate an ensemble of pipelines (aka campaign) since each pipeline progresses at different rates of progress; (ii) diverse resource management and scheduling requirements, different parts of a workflow might have different resource requirements; (iii) fault-tolerance of the tools and executing processes. Thus, we used RADICAL-EnTK (Ensemble Toolkit), as part of the ExaWorks SDK, to build corresponding workflow applications. It allowed us to utilize 8000 compute nodes on Frontier at OLCF/ORNL ($85\%$ of all Frontier's nodes) and to make 7875 simulation runs while having 1000 concurrent simulation runs at each moment. The total runtime was 3.3 hours, and we achieved $90\%$ of the resource utilization (allocated resources: 448,000 CPU cores and 64,000 GPUs).
%------------------------------------------
\section{Discussion}
% \jhanote{Rafael: Please discuss: (1) WCI and (2) SWAS)}

Addressing the implications of the innocuous statement, ``Workflows are the new applications''~\citep{ben2020workflows} will not be easy, nor obvious.  ExaWorks provides the first targeted and community-driven project that delivered building blocks for workflow applications and systems.
Notably, the ExaWorks experience reiterates that both social ``structures" (i.e., for PSI/J) and technical ``open building blocks" (integration of diverse components) are needed. It proves that one does not necessarily have to spin one's end-to-end and closed workflow
system. ExaWorks has spearheaded community efforts such as PSI/J, which is
becoming a widely accepted and supported interface for batch-queue systems.

The Workflows Community Initiative (WCI), driven by the ExaWorks Principal Investigators and a collective of global workflow researchers, is a voluntary effort to consolidate the workflows community. This initiative encompasses users, developers, researchers, and facility representatives, offering them a platform to access community resources and capabilities. The purpose is to facilitate the discovery of software products, related endeavors, events, technical reports, and more, fostering community-wide engagement to address the significant workflow challenges. Through a community-focused strategy, in tight collaboration with WCI, ExaWorks has successfully hosted four Workflow Summits. These events have drawn the participation of hundreds of researchers, developers, and facility delegates. The principal achievements of these summits include the production of technical reports~\citep{wcs2022, wcs2024} that summarize the discussions held, with a pivotal outcome being the formulation of a community roadmap for workflow research and development. ExaWorks and WCI have established themselves as community frontrunners, promoting ``workflow thinking'' and offering comprehensive hands-on training to an extensive user base.

With the successful conclusion of the ExaWorks project, our collective efforts are now transitioning towards the SWAS (Center for Stewardship and Advancement of Workflows and Application Services) project~\citep{swas-web}. The SWAS initiative is dedicated to the stewardship and advancement of workflow tools, components, and application services, addressing the ever-growing needs of the scientific workflow community. This project represents a strategic pivot towards harnessing and enhancing the substantial groundwork laid by ExaWorks, aiming to create a more integrated, efficient, and sustainable ecosystem for workflow and application services. By emphasizing a community-centric approach, SWAS seeks to foster innovation, facilitate collaboration, and streamline the adoption of advanced workflow solutions across diverse scientific domains.

The proposed activities under SWAS are meticulously designed to build upon and expand ExaWorks' achievements. Central to SWAS's mission is establishing a comprehensive Workflows SDK, which will encapsulate common interfaces, abstractions, and testing infrastructures, simplifying workflow tool integration and deployment across varied computing environments. Moreover, SWAS commits to developing end-to-end cross-site workflow testing services and user experience (UX) research, aiming to elevate the usability and reliability of workflow tools. Through these endeavors, SWAS intends to cultivate a vibrant ecosystem that supports workflow software's lifecycle and actively engages with the community through training, outreach, and partnership initiatives. By drawing on the collective expertise and infrastructure developed during the ExaWorks project, SWAS is poised to significantly advance the state of scientific workflow management, fostering an environment where research, development, education, and training coalesce to propel scientific discovery forward.
%------------------------------------------
\section*{Acknowledgments}
This research was supported by the Exascale Computing Project (17-SC-20-SC), a collaborative effort of the U.S. Department of Energy Office of Science and the National Nuclear Security Administration. This work was performed under the auspices of the U.S. Department of Energy by Lawrence Livermore National Laboratory under Contract DE-AC52-07NA27344 (LLNL-CONF-826133), Argonne National Laboratory under Contract DE-AC02-06CH11357, and Brookhaven National Laboratory under Contract DESC0012704. This research used resources of the OLCF at ORNL, supported by the Office of Science of the U.S. DOE under Contract No. DE-AC05-00OR22725. 

%Bibliography
\bibliographystyle{unsrtnat}
\bibliography{references}

\begin{thebibliography}{35}
\providecommand{\natexlab}[1]{#1}
\providecommand{\url}[1]{\texttt{#1}}
\expandafter\ifx\csname urlstyle\endcsname\relax
  \providecommand{\doi}[1]{doi: #1}\else
  \providecommand{\doi}{doi: \begingroup \urlstyle{rm}\Url}\fi

\bibitem[Ferreira~da Silva et~al.(2021{\natexlab{a}})Ferreira~da Silva, Casanova, Chard, Altintas, Badia, Balis, Coleman, Coppens, Di~Natale, Enders, et~al.]{da2021community}
Rafael Ferreira~da Silva, Henri Casanova, Kyle Chard, Ilkay Altintas, Rosa~M Badia, Bartosz Balis, Taina Coleman, Frederik Coppens, Frank Di~Natale, Bjoern Enders, et~al.
\newblock A community roadmap for scientific workflows research and development.
\newblock In \emph{2021 IEEE Workshop on Workflows in Support of Large-Scale Science (WORKS)}, pages 81--90. IEEE, 2021{\natexlab{a}}.

\bibitem[Badia~Sala et~al.(2017)Badia~Sala, Ayguad{\'e}~Parra, and Labarta~Mancho]{badia2017workflows}
Rosa~Maria Badia~Sala, Eduard Ayguad{\'e}~Parra, and Jes{\'u}s~Jos{\'e} Labarta~Mancho.
\newblock Workflows for science: A challenge when facing the convergence of hpc and big data.
\newblock \emph{Supercomputing frontiers and innovations}, 4\penalty0 (1):\penalty0 27--47, 2017.

\bibitem[Ferreira~da Silva et~al.(2024{\natexlab{a}})Ferreira~da Silva, Badia, Bard, Foster, Jha, and Suter]{dasilva2024computer}
Rafael Ferreira~da Silva, Rosa~M. Badia, Deborah Bard, Ian~T. Foster, Shantenu Jha, and Frédéric Suter.
\newblock Frontiers in scientific workflows: Pervasive integration with hpc.
\newblock \emph{IEEE Computer}, 57\penalty0 (8), 2024{\natexlab{a}}.
\newblock \doi{10.1109/MC.2024.3401542}.

\bibitem[Ferreira~da Silva et~al.(2023)Ferreira~da Silva, Badia, Bala, Bard, Bremer, Buckley, Caino-Lores, Chard, Goble, Jha, Katz, Laney, Parashar, Suter, Tyler, Uram, Altintas, et~al.]{wcs2022}
Rafael Ferreira~da Silva, Rosa~M. Badia, Venkat Bala, Debbie Bard, Timo Bremer, Ian Buckley, Silvina Caino-Lores, Kyle Chard, Carole Goble, Shantenu Jha, Daniel~S. Katz, Daniel Laney, Manish Parashar, Frederic Suter, Nick Tyler, Thomas Uram, Ilkay Altintas, et~al.
\newblock {Workflows Community Summit 2022: A Roadmap Revolution}.
\newblock Technical Report ORNL/TM\%-2023/2885, Oak Ridge National Laboratory, March 2023.

\bibitem[Saadi et~al.(2021)Saadi, Alfe, Babuji, Bhati, Blaiszik, Brettin, Chard, Chard, Coveney, Trifan, Brace, Clyde, Foster, Gibbs, Jha, Keipert, Kurth, Kranzlmüller, Lee, Li, Ma, Merzky, Mathias, Partin, Yin, Ramanathan, Shah, Stern, Stevens, Tan, Titov, Tsaris, Turilli, Dam, Wan, and Wifling]{saadi2021impeccable}
Aymen~Al Saadi, Dario Alfe, Yadu Babuji, Agastya Bhati, Ben Blaiszik, Thomas Brettin, Kyle Chard, Ryan Chard, Peter Coveney, Anda Trifan, Alex Brace, Austin Clyde, Ian Foster, Tom Gibbs, Shantenu Jha, Kristopher Keipert, Thorsten Kurth, Dieter Kranzlmüller, Hyungro Lee, Zhuozhao Li, Heng Ma, Andre Merzky, Gerald Mathias, Alexander Partin, Junqi Yin, Arvind Ramanathan, Ashka Shah, Abraham Stern, Rick Stevens, Li~Tan, Mikhail Titov, Aristeidis Tsaris, Matteo Turilli, Huub~Van Dam, Shunzhou Wan, and David Wifling.
\newblock Impeccable: Integrated modeling pipeline for covid cure by assessing better leads.
\newblock \emph{50th International Conference on Parallel Processing (ICPP ’21), August 9–12, 2021, Lemont, IL, USA. ACM, New York, NY, USA, 12 pages}, 2021.
\newblock \doi{\url{https://doi.org/10.1145/3472456.3473524}}.

\bibitem[exa(2021)]{exaworks-survey}
{ExaWorks Workflow Survey}.
\newblock Technical report, Exascale Computing Project, 2021.
\newblock \url{https://exaworks.org/documents/ExaWorksSurveyReport.pdf}.

\bibitem[git(2024)]{github-workflows}
Awesome workflow engines.
\newblock \url{https://github.com/meirwah/awesome-workflow-engines}, 2024.

\bibitem[Al-Saadi et~al.(2021)Al-Saadi, Ahn, Babuji, Chard, Corbett, Hategan, Herbein, Jha, Laney, Merzky, et~al.]{al2021exaworks}
Aymen Al-Saadi, Dong~H Ahn, Yadu Babuji, Kyle Chard, James Corbett, Mihael Hategan, Stephen Herbein, Shantenu Jha, Daniel Laney, Andre Merzky, et~al.
\newblock Exaworks: Workflows for exascale.
\newblock In \emph{2021 IEEE Workshop on Workflows in Support of Large-Scale Science (WORKS)}, pages 50--57. IEEE, 2021.

\bibitem[Ferreira~da Silva et~al.(2021{\natexlab{b}})Ferreira~da Silva, Casanova, Chard, Altintas, Badia, Balis, Coleman, Coppens, Di~Natale, Enders, Fahringer, Filgueira, et~al.]{ferreiradasilva2021works}
Rafael Ferreira~da Silva, Henri Casanova, Kyle Chard, Ilkay Altintas, Rosa~M Badia, Bartosz Balis, Tain\~a Coleman, Frederik Coppens, Frank Di~Natale, Bjoern Enders, Thomas Fahringer, Rosa Filgueira, et~al.
\newblock A community roadmap for scientific workflows research and development.
\newblock In \emph{2021 IEEE Workshop on Workflows in Support of Large-Scale Science (WORKS)}, pages 81--90, 2021{\natexlab{b}}.
\newblock \doi{10.1109/WORKS54523.2021.00016}.

\bibitem[Turilli et~al.(2019)Turilli, Balasubramanian, Merzky, Paraskevakos, and Jha]{turilli2019middleware}
Matteo Turilli, Vivek Balasubramanian, Andre Merzky, Ioannis Paraskevakos, and Shantenu Jha.
\newblock Middleware building blocks for workflow systems.
\newblock \emph{Computing in Science \& Engineering}, 21\penalty0 (4):\penalty0 62--75, 2019.

\bibitem[Hategan-Marandiuc et~al.(2023)Hategan-Marandiuc, Merzky, Collier, Maheshwari, Ozik, Turilli, Wilke, Wozniak, Chard, Foster, et~al.]{hategan2023psi}
Mihael Hategan-Marandiuc, Andre Merzky, Nicholson Collier, Ketan Maheshwari, Jonathan Ozik, Matteo Turilli, Andreas Wilke, Justin~M. Wozniak, Kyle Chard, Ian Foster, et~al.
\newblock Psi/j: A portable interface for submitting, monitoring, and managing jobs.
\newblock In \emph{2023 IEEE 19th International Conference on e-Science (e-Science)}, pages 1--10. IEEE, 2023.

\bibitem[Turner et~al.(2022)Turner, Belak, Barton, Bement, Carlson, Carson, DeWitt, Fattebert, Hodge, Jibben, et~al.]{turner2022exaam}
John~A Turner, James Belak, Nathan Barton, Matthew Bement, Neil Carlson, Robert Carson, Stephen DeWitt, Jean-Luc Fattebert, Neil Hodge, Zechariah Jibben, et~al.
\newblock Exaam: Metal additive manufacturing simulation at the fidelity of the microstructure.
\newblock \emph{The International Journal of High Performance Computing Applications}, 36\penalty0 (1):\penalty0 13--39, 2022.

\bibitem[Wozniak(2017)]{Workflow_Survey_2017}
Justin~M. Wozniak.
\newblock {{ECP}} workflow survey report.
\newblock Technical Report ANL-24/20, Argonne National Laboratory, 2017.

\bibitem[Ahn et~al.(2014)Ahn, Garlick, Grondona, Lipari, Springmeyer, and Schulz]{flux}
Dong~H. Ahn, Jim Garlick, Mark Grondona, Don Lipari, Becky Springmeyer, and Martin Schulz.
\newblock Flux: A next-generation resource management framework for large hpc centers.
\newblock In \emph{2014 43rd International Conference on Parallel Processing Workshops}, pages 9--17, 2014.
\newblock \doi{10.1109/ICPPW.2014.15}.

\bibitem[Ahn et~al.(2020)Ahn, Bass, Chu, Garlick, Grondona, Herbein, Ingólfsson, Koning, Patki, Scogland, Springmeyer, and Taufer]{AHN2020202}
Dong~H. Ahn, Ned Bass, Albert Chu, Jim Garlick, Mark Grondona, Stephen Herbein, Helgi~I. Ingólfsson, Joseph Koning, Tapasya Patki, Thomas~R.W. Scogland, Becky Springmeyer, and Michela Taufer.
\newblock Flux: Overcoming scheduling challenges for exascale workflows.
\newblock \emph{Future Generation Computer Systems}, 110:\penalty0 202--213, 2020.
\newblock ISSN 0167-739X.
\newblock \doi{https://doi.org/10.1016/j.future.2020.04.006}.
\newblock URL \url{https://www.sciencedirect.com/science/article/pii/S0167739X19317169}.

\bibitem[Di~Natale et~al.(2019)Di~Natale, Bhatia, Carpenter, Neale, Kokkila-Schumacher, Oppelstrup, Stanton, Zhang, Sundram, Scogland, Dharuman, Surh, Yang, Misale, Schneidenbach, Costa, Kim, D'Amora, Gnanakaran, Nissley, Streitz, Lightstone, Bremer, Glosli, and Ing\'{o}lfsson]{dinatale19maestro}
Francesco Di~Natale, Harsh Bhatia, Timothy~S. Carpenter, Chris Neale, Sara Kokkila-Schumacher, Tomas Oppelstrup, Liam Stanton, Xiaohua Zhang, Shiv Sundram, Thomas R.~W. Scogland, Gautham Dharuman, Michael~P. Surh, Yue Yang, Claudia Misale, Lars Schneidenbach, Carlos Costa, Changhoan Kim, Bruce D'Amora, Sandrasegaram Gnanakaran, Dwight~V. Nissley, Fred Streitz, Felice~C. Lightstone, Peer-Timo Bremer, James~N. Glosli, and Helgi~I. Ing\'{o}lfsson.
\newblock A massively parallel infrastructure for adaptive multiscale simulations: Modeling {RAS} initiation pathway for cancer.
\newblock In \emph{Proceedings of the International Conference for High Performance Computing, Networking, Storage and Analysis}, pages 1--16, 2019.
\newblock \doi{10.1145/3295500.3356197}.

\bibitem[Babuji et~al.(2019)Babuji, Woodard, Li, Katz, Clifford, Kumar, Lacinski, Chard, Wozniak, Foster, Wilde, and Chard]{babuji19parsl}
Yadu Babuji, Anna Woodard, Zhuozhao Li, Daniel~S. Katz, Ben Clifford, Rohan Kumar, Luksaz Lacinski, Ryan Chard, Justin~M. Wozniak, Ian Foster, Michael Wilde, and Kyle Chard.
\newblock Parsl: Pervasive parallel programming in {P}ython.
\newblock In \emph{28th ACM International Symposium on High-Performance Parallel and Distributed Computing (HPDC)}, pages 25--36, 2019.
\newblock \doi{10.1145/3307681.3325400}.

\bibitem[Balasubramanian et~al.(2016)Balasubramanian, Treikalis, Weidner, and Jha]{balasubramanian2016ensemble}
Vivekanandan Balasubramanian, Antons Treikalis, Ole Weidner, and Shantenu Jha.
\newblock Ensemble toolkit: Scalable and flexible execution of ensembles of tasks.
\newblock In \emph{2016 45th International Conference on Parallel Processing (ICPP)}, pages 458--463. IEEE, 2016.

\bibitem[Merzky et~al.(2021)Merzky, Turilli, Titov, Al-Saadi, and Jha]{merzky2021design}
Andre Merzky, Matteo Turilli, Mikhail Titov, Aymen Al-Saadi, and Shantenu Jha.
\newblock Design and performance characterization of {Radical-Pilot} on leadership-class platforms.
\newblock \emph{IEEE Transactions on Parallel and Distributed Systems}, 33\penalty0 (4):\penalty0 818--829, 2021.

\bibitem[Partee et~al.(2022)Partee, Ellis, Rigazzi, Shao, Bachman, Marques, and Robbins]{smartsim}
Sam Partee, Matthew Ellis, Alessandro Rigazzi, Andrew~E. Shao, Scott Bachman, Gustavo Marques, and Benjamin Robbins.
\newblock Using machine learning at scale in numerical simulations with {SmartSim}: {A}n application to ocean climate modeling.
\newblock \emph{Journal of Computational Science}, 62:\penalty0 101707, 2022.
\newblock ISSN 1877-7503.
\newblock \doi{https://doi.org/10.1016/j.jocs.2022.101707}.
\newblock URL \url{https://www.sciencedirect.com/science/article/pii/S1877750322001065}.

\bibitem[Wozniak et~al.(2013)Wozniak, Armstrong, Wilde, Katz, Lusk, and Foster]{wozniak2013swift}
Justin~M Wozniak, Timothy~G Armstrong, Michael Wilde, Daniel~S Katz, Ewing Lusk, and Ian~T Foster.
\newblock {Swift/T}: {L}arge-scale application composition via distributed-memory dataflow processing.
\newblock In \emph{2013 13th IEEE/ACM International Symposium on Cluster, Cloud, and Grid Computing}, pages 95--102. IEEE, 2013.

\bibitem[{The Vue.js Team}(2024)]{vuejs-web}
{The Vue.js Team}.
\newblock Vue.js, 2024.
\newblock URL \url{https://vuejs.org/}.

\bibitem[{The ExaWorks Project}(2024{\natexlab{a}})]{sdk-readthedocs}
{The ExaWorks Project}.
\newblock {ExaWorks}: {Software Development Kit}, 2024{\natexlab{a}}.
\newblock URL \url{https://exaworkssdk.readthedocs.io/en/latest/}.

\bibitem[{The ExaWorks Project}(2024{\natexlab{b}})]{sdk-container}
{The ExaWorks Project}.
\newblock {ExaWorks Software Development Kit Docker Container}, 2024{\natexlab{b}}.
\newblock URL \url{https://github.com/ExaWorks/SDK/tree/master/docker}.

\bibitem[{The Binder Project}(2024)]{binder-web}
{The Binder Project}.
\newblock Reproducible, sharable, open, interactive computing environments, 2024.
\newblock URL \url{https://mybinder.org/}.

\bibitem[Ward et~al.(2021)Ward, Sivaraman, Pauloski, Babuji, Chard, Dandu, Redfern, Assary, Chard, Curtiss, Thakur, and Foster]{wardColmena2021}
L.~Ward, G.~Sivaraman, J.~Pauloski, Y.~Babuji, R.~Chard, N.~Dandu, P.~C. Redfern, R.~S. Assary, K.~Chard, L.~A. Curtiss, R.~Thakur, and I.~Foster.
\newblock Colmena: Scalable machine-learning-based steering of ensemble simulations for high performance computing.
\newblock In \emph{2021 IEEE/ACM Workshop on Machine Learning in High Performance Computing Environments (MLHPC)}, pages 9--20, Los Alamitos, CA, USA, nov 2021. IEEE Computer Society.
\newblock \doi{10.1109/MLHPC54614.2021.00007}.
\newblock URL \url{https://doi.ieeecomputersociety.org/10.1109/MLHPC54614.2021.00007}.

\bibitem[Alsaadi et~al.(2022)Alsaadi, Ward, Merzky, Chard, Foster, Jha, and Turilli]{alsaadiParslRP2022}
A.~Alsaadi, L.~Ward, A.~Merzky, K.~Chard, I.~Foster, S.~Jha, and M.~Turilli.
\newblock {RADICAL-Pilot} and {Parsl}: Executing heterogeneous workflows on {HPC} platforms.
\newblock In \emph{2022 IEEE/ACM Workshop on Workflows in Support of Large-Scale Science (WORKS)}, pages 27--34, Los Alamitos, CA, USA, nov 2022. IEEE Computer Society.
\newblock \doi{10.1109/WORKS56498.2022.00009}.
\newblock URL \url{https://doi.ieeecomputersociety.org/10.1109/WORKS56498.2022.00009}.

\bibitem[den Burger et~al.(2010)den Burger, Jacobs, Kielmann, Merzky, Weidner, and Kaiser]{den2010price}
Mathijs den Burger, Ceriel Jacobs, Thilo Kielmann, Andre Merzky, Ole Weidner, and Hartmut Kaiser.
\newblock What is the price of simplicity? a cross-platform evaluation of the saga api.
\newblock In \emph{Euro-Par 2010-Parallel Processing: 16th International Euro-Par Conference, Ischia, Italy, August 31-September 3, 2010, Proceedings, Part I 16}, pages 392--404. Springer, 2010.

\bibitem[{The ExaWorks Team}(2024)]{psijspec}
{The ExaWorks Team}.
\newblock {A Portable Submission Interface for Jobs (PSI/J)}.
\newblock \url{https://purl.org/psij.io/spec}, 2024.

\bibitem[Engelmann et~al.(2022)Engelmann, Kuchar, Boehm, Brim, Naughton, Somnath, Atchley, Lange, Mintz, and Arenholz]{engelmann22intersect}
Christian Engelmann, Olga Kuchar, Swen Boehm, Michael~J. Brim, Thomas Naughton, Suhas Somnath, Scott Atchley, Jack Lange, Ben Mintz, and Elke Arenholz.
\newblock The {INTERSECT} open federated architecture for the laboratory of the future.
\newblock In \emph{Communications in Computer and Information Science (CCIS): Accelerating Science and Engineering Discoveries Through Integrated Research Infrastructure for Experiment, Big Data, Modeling and Simulation. \url{https://smc.ornl.gov} Smoky Mountains Computational Sciences \% Engineering Conference (SMC) 2022}, volume 1690, pages 173--190. Springer, Cham, aug 2022.
\newblock ISBN 978-3-031-23605-1.
\newblock \doi{10.1007/978-3-031-23606-8_11}.

\bibitem[Chow et~al.(2023)Chow, Zvyagin, Brace, Hippe, Deng, Zhang, Bohorquez, Clyde, Kale, Perez-Rivera, Ma, Mann, Irvin, Ozgulbas, Vassilieva, Pauloski, Ward, Hayot-Sasson, Emani, Foreman, Xie, Lin, Shukla, Nie, Romero, Dallago, Vahdat, Xiao, Gibbs, Foster, Davis, Papka, Brettin, Stevens, Anandkumar, Vishwanath, and Ramanathan]{chow23genslms}
Edmond Chow, Maxim Zvyagin, Alexander Brace, Kyle Hippe, Yuntian Deng, Bin Zhang, Cindy~Orozco Bohorquez, Austin Clyde, Bharat Kale, Danilo Perez-Rivera, Heng Ma, Carla~M. Mann, Michael Irvin, Defne~G. Ozgulbas, Natalia Vassilieva, James~Gregory Pauloski, Logan Ward, Valerie Hayot-Sasson, Murali Emani, Sam Foreman, Zhen Xie, Diangen Lin, Maulik Shukla, Weili Nie, Josh Romero, Christian Dallago, Arash Vahdat, Chaowei Xiao, Thomas Gibbs, Ian Foster, James~J. Davis, Michael~E. Papka, Thomas Brettin, Rick Stevens, Anima Anandkumar, Venkatram Vishwanath, and Arvind Ramanathan.
\newblock Genslms: Genome-scale language models reveal sars-cov-2 evolutionary dynamics.
\newblock \emph{Int. J. High Perform. Comput. Appl.}, 37\penalty0 (6):\penalty0 683--705, nov 2023.
\newblock \doi{10.1177/10943420231201154}.

\bibitem[Ward et~al.(2025)Ward, Pauloski, Hayot-Sasson, Babuji, Brace, Chard, Chard, Thakur, and Foster]{colmenaecp}
Logan Ward, Gregory Pauloski, Valerie Hayot-Sasson, Yadu Babuji, Alexander Brace, Ryan Chard, Kyle Chard, Rajeev Thakur, and Ian Foster.
\newblock Employing artificial intelligence to steer exascale workflows with colmena.
\newblock \emph{Submitted to IJHPCA}, 2025.

\bibitem[Ben-Nun et~al.(2020)Ben-Nun, Gamblin, Hollman, Krishnan, and Newburn]{ben2020workflows}
Tal Ben-Nun, Todd Gamblin, Daisy~S Hollman, Hari Krishnan, and Chris~J Newburn.
\newblock Workflows are the new applications: Challenges in performance, portability, and productivity.
\newblock In \emph{2020 IEEE/ACM International Workshop on Performance, Portability and Productivity in HPC (P3HPC)}, pages 57--69. IEEE, 2020.

\bibitem[Ferreira~da Silva et~al.(2024{\natexlab{b}})Ferreira~da Silva, Bard, Chard, DeWitt, Foster, Gibbs, Goble, Godoy, Gustafsson, Haus, Hudson, Jha, Los, Paine, Suter, Ward, Wilkinson, et~al.]{wcs2024}
Rafael Ferreira~da Silva, Deborah Bard, Kyle Chard, Shaun DeWitt, Ian~T. Foster, Tom Gibbs, Carole Goble, William Godoy, Johan Gustafsson, Utz\%-Uwe Haus, Stephen Hudson, Shantenu Jha, Laila Los, Drew Paine, Frederic Suter, Logan Ward, Sean Wilkinson, et~al.
\newblock {Workflows Community Summit 2024 Future Trends and Challenges in Scientific Workflows}.
\newblock Technical Report ORNL/TM\%-2024/3573, Oak Ridge National Laboratory, 2024{\natexlab{b}}.

\bibitem[{The SWAS Project}(2024)]{swas-web}
{The SWAS Project}.
\newblock {SWAS}: Sustaining workflows \& application services, 2024.
\newblock URL \url{https://swas.center/}.

\end{thebibliography}

\end{document}